\documentclass[fleqn,twoside]{article}
\usepackage[headings]{espcrc2}

\readRCS
$Id: espcrc2.tex,v 1.2 2004/02/24 11:22:11 spepping Exp $
\usepackage{graphicx}
\usepackage[figuresright]{rotating}

\newcommand{\AmS}{{\protect\the\textfont2
  A\kern-.1667em\lower.5ex\hbox{M}\kern-.125emS}}

\newcommand{\be}{\begin{equation}}
\newcommand{\ee}{\end{equation}}
\newcommand{\bea}{\begin{eqnarray}}
\newcommand{\eea}{\end{eqnarray}}
\newcommand{\beq}{\begin{equation}}
\newcommand{\eeq}{\end{equation}}

\title{An Automatic System to Discriminate Malignant from Benign Massive Lesions on Mammograms}

\author{A. Retico\address[INFN]{Istituto Nazionale di Fisica Nucleare, Largo Pontecorvo 3, 56127 Pisa, Italy }\thanks{Corresponding author.  {\em E-mail address:} alessandra.retico@df.unipi.it (A. Retico).
{\em Tel:} +39 0502214459; {\em fax:} +39 0502214317.},
        P. Delogu\addressmark\address[DipPhys]{Dipartimento di Fisica dell'Universit\`a di Pisa, Largo Pontecorvo 3, 56127 Pisa, Italy},
	M.E. Fantacci\addressmark[INFN]\addressmark[DipPhys],
	P. Kasae\address{The Abdus Salam International Center for Theoretical Physics, Strada Costiera 11, P.O. Box 563, I-34100 Trieste, Italy}}

\runtitle{An Automatic System to Discriminate Malignant from Benign Massive Lesions on Mammograms}
\runauthor{A. Retico et al.}

\begin{document}

\begin{abstract}
Mammography is widely recognized as the most reliable technique for early detection of breast cancers. Automated or semi-automated computerized classification schemes can be very useful in assisting  radiologists  with a second opinion about the visual diagnosis of breast lesions, thus leading to a reduction in the number of unnecessary biopsies.
We present a computer-aided diagnosis (CADi) system for the characterization of  massive lesions in mammograms, whose aim is to distinguish malignant from benign masses.
The CADi system  we realized is based on a three-stage algorithm: a) a segmentation technique extracts the contours of the massive lesion from the image; b) sixteen features based on size and shape of the lesion are computed; c) a neural classifier merges the features into an estimated likelihood of malignancy. 
A dataset of 226 massive lesions (109 malignant and 117 benign) 
has been used in this study. 
The system performances have been evaluated terms of the receiver-operating characteristic (ROC) analysis, obtaining $A_z = 0.80\pm 0.04$ as the estimated area under the ROC curve.

\vspace{1pc}


\vspace{1pc}

\end{abstract}

\maketitle

\section*{Introduction}

Breast cancer is still one of the main causes of death among women,
despite early detections have recently contributed to a significant
decrease in the breast-cancer
mortality~\cite{Greenlee,Levi14}.  Mammography is an
effective technique for detecting breast cancer in its early
stages~\cite{Zuckerman}.  Once a massive lesion is detected on a
mammogram, the radiologist recommends further investigations,
depending on the likelihood of malignancy he assigns to the lesion.
However, the characterization of massive lesions merely on the basis
of a visual analysis of the mammogram is a very difficult task and a
high number of unnecessary biopsies are actually performed in the
routine clinical activity.  The rate of positive findings for cancers
at biopsy ranges from 15\% to 30\%~\cite{Adler}, i.e.  the specificity
in differentiating malignant from benign lesions merely on the basis of 
the radiologist's interpretation of mammograms is rather low.  
Methods to improve mammographic specificity without missing
cancer have to be developed. 
Computerized
method have recently shown a great potential in assisting radiologists in the visual diagnosis of the lesions, by providing them with a second
opinion about the degree of malignancy of a lesion~\cite{Huo1,Huo2,Sahiner0}.

The computerized system for the classification of benign and malignant
massive lesions we describe in this paper is a semi-automated one,
i.e. it provides a likelihood of malignancy for a physician-selected
region of a mammogram.

This paper is structured as follows: a technical description of the method is given in sec.~\ref{sec:method}; section~\ref{sec:dataset} illustrates the dataset of mammograms we used for this study and sec.~\ref{sec:results} reports on the performances the CADi system achieved in differentiating malignant from benign massive lesions.

\section{Description of the CADi system}
\label{sec:method}
The system for characterizing massive lesions
we realized is based on a three-stage algorithm: 
first, a segmentation technique extracts the massive lesion from the image; then, several features based on size, shape and texture 
of the lesion are computed; 
finally,  a neural classifier merges the features into a likelihood of
 malignancy for that lesion.

\subsection{Segmentation}

Massive lesions are extremely variable in size, shape and density; they 
can exhibit a
very poor image contrast or can be highly connected to the surrounding
parenchymal tissue. For these reasons segmenting massive lesions from the 
nonuniform normal breast tissue is considered  a 
 nontrivial task and much efforts have already gone
through this issue~\cite{Wirth,Amini,Sahiner,Matthew}.

\begin{figure}[htb]
\vspace{9pt}
\begin{center}
\includegraphics[width=6.0cm]{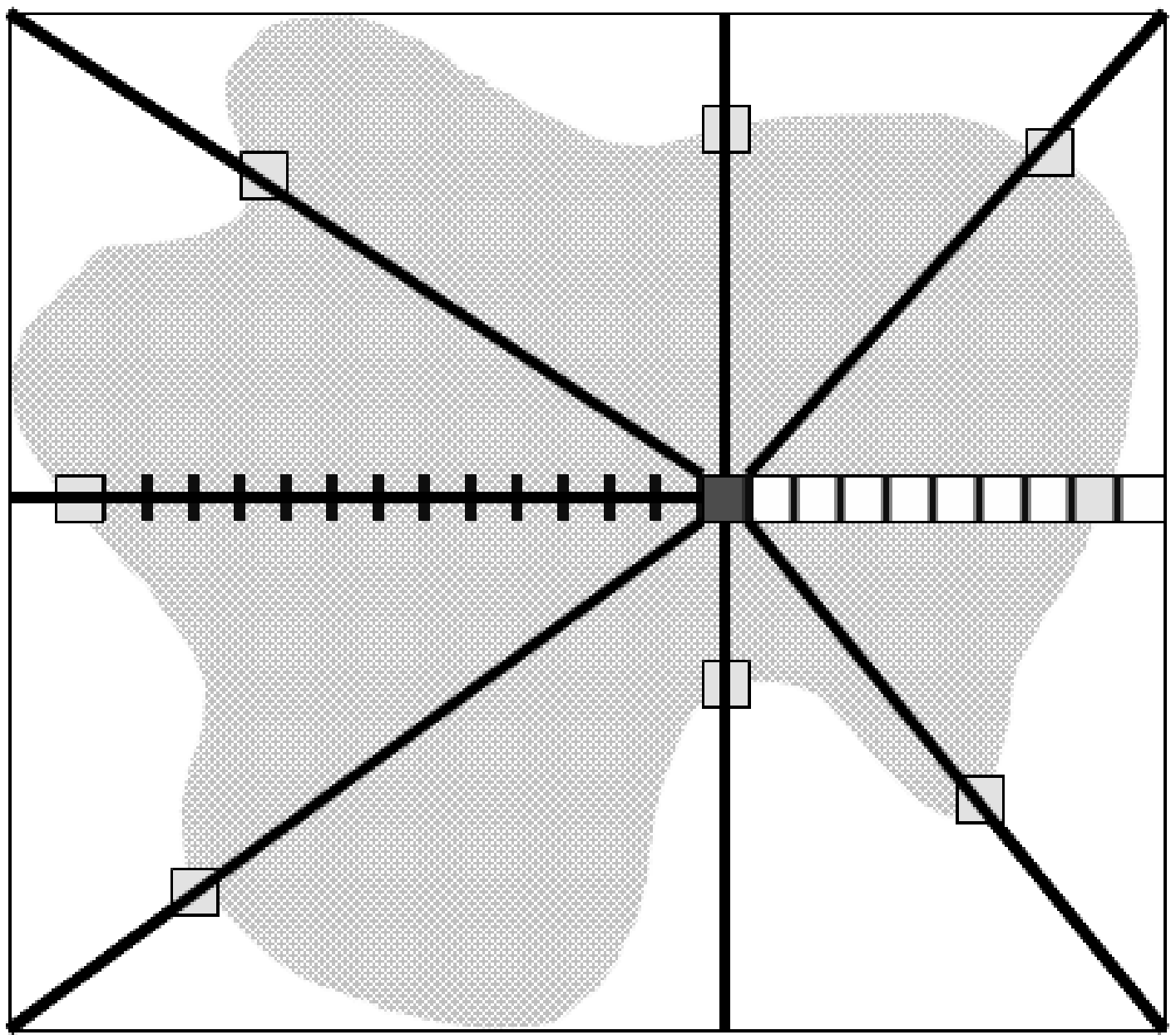}
\includegraphics[width=5.0cm]{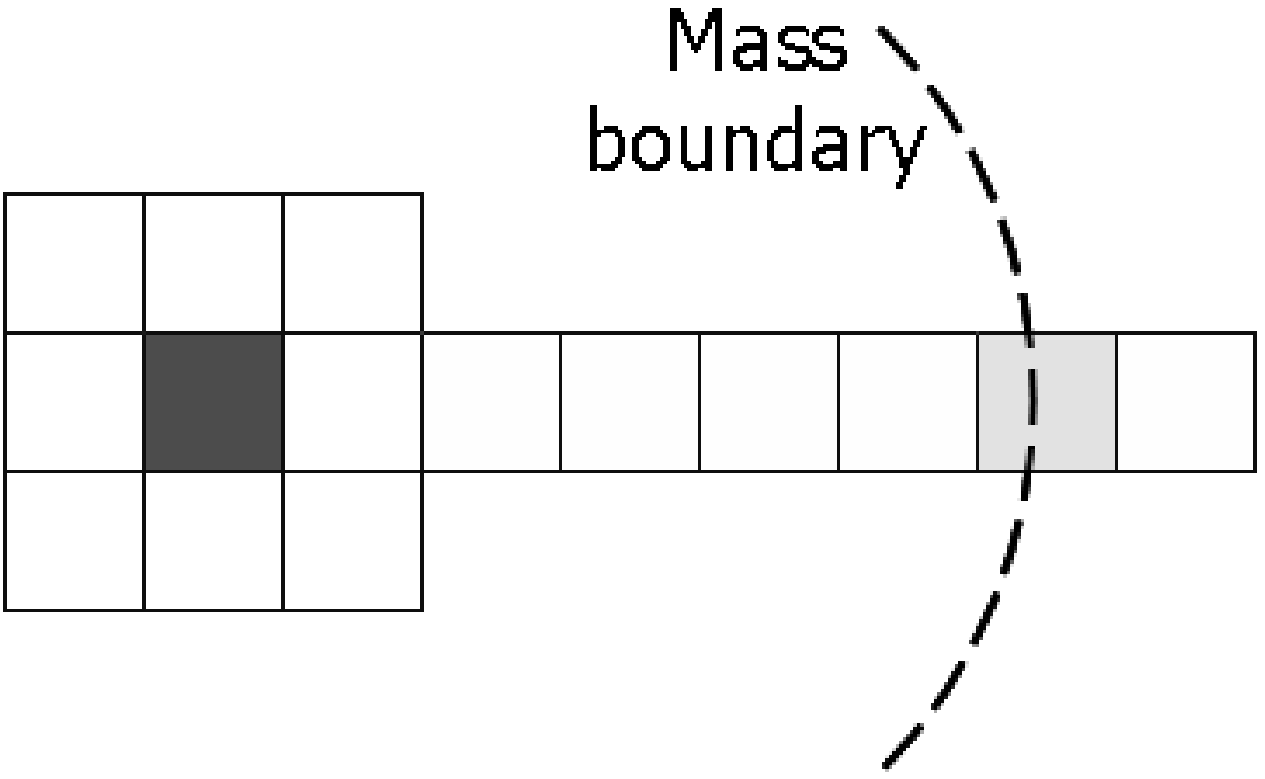}
\end{center}
\caption{Basic schema of the segmentation procedure.}
\label{fig:segm}
\end{figure}

The segmentation algorithm we developed is an extension and a
refinement of the strategy proposed in~\cite{Chen} for the
mass segmentation in the computerized analysis of breast tumors on
sonograms.
A massive lesion is automatically identified within a rectangular Region Of
Interest (ROI) interactively chosen by the radiologist. 
The ROIs contain
the lesions as well as a considerable part of normal tissue. 
In our segmentation procedure the non-tumor 
regions  in a ROI are removed by applying the following processing
steps (fig.~\ref{fig:segm}): 
\begin{itemize}
\item
The pixel characterized by the maximum-intensity value in the central area of the ROI is taken as the seed point for the segmentation algorithm. 
\item
A number of radial lines are depicted from the  seed point
to the boundary of the ROI.
\item
For each pixel along each radial line the local variance (i.e. the variance of the entries of a $n \times n$ matrix containing the pixel and its neighborhood)  is computed. 
The pixel maximizing the local variance  is most likely the one located on the
boundary between the mass and the surrounding tissue and it is referred as critical point.
\item
 The critical points determined for each radial line are linearly interpolated and a coarse boundary for the lesion is determined.
\item
The pixels inside  this coarse region are taken as 
new seed points for iterating the procedure in order to end up with  
a more accurate identification of the shape
of the lesion.
\item
A set of candidates to represent the mass boundary is thus obtained. Every identified point is accepted and the area inside  the resulting thick border is filled. 
Once the possibly present non-connected objects are removed, the thick boundary and the area inside it are accepted as the segmented massive lesion. 
\end{itemize}

\subsection{Feature extraction}

Once the masses have been segmented out of the surrounding normal
tissue, a set of suitable features are computed in order to
allow a decision-making system to distinguish benign from malignant
lesions~\cite{Christoyianni,Qian,Hadjiiski,Sahiner2}.
The degree of malignancy of a lesion is generally correlated to the 
appearance of arms and spiculations  on the mass boundary. 
The more irregular the mass shape, 
the higher the degree of malignancy possibly associated to that lesion. 
Our CADi
system extracts 16 features from the segmented lesions:
the {\em mass area} $A$; 
the {\em mass perimeter} $P$;  
the {\em circularity} $C ={4\pi A}/{P^2}$; 
the {\em mean} and the {\em standard deviation} of the {\em  normalized radial length} (i.e. the Euclidean distance from the center of mass of the segmented
lesion to the $i^{\rm th}$ pixel on the perimeter and normalized  to the maximum distance for that mass);  
the {\em radial length entropy} (i.e. a probabilistic measure computed from the histogram of the normalized radial length as $E = - \sum_{k=1}^{N_{\rm bins}} P_k \log{P_k}$, where $P_k$ is the probability that the normalized radial length is between $d(i)$ and $d(i) + 1/N_{\rm bins}$ and $N_{\rm bins}$ is the  number of bins the normalized histogram has been divided in); 
the {\em zero crossing} (i.e. a count of the number of times the radial distance plot crosses the average radial distance); 
the {\em maximum} and the {\em minimum axis} of the lesion; 
the {\em mean} and the  {\em standard deviation} of the {\em variation ratio} (i.e. the modulus of the variations of the radial lengths from their mean value are computed and only those exceeding the value $var_{\rm max}/2 $, where $var_{\rm max}$ is the maximum variation, are considered as dominant variations and averaged);
the {\em convexity} (i.e. the ratio between the mass area and the area 
of the smallest convex containing the mass); 
the {\em mean}, the  {\em standard deviation}, the {\em skewness} and the {\em kurtosis} of the {\em mass grey-level intensity values}.

The 16 features were chosen with the aim of enlightening the 
spiculation characteristics of the
lesions.  The first 12 features in the
above description are related to the mass shape and 
have some evident correlations with the degree of
spiculation of the lesions. Nevertheless,  the remaining 4 features
derived from the grey-level intensity distribution of the segmented
area also aim at investigating the degree of mass spiculation: the standard
deviation, the skewness and the kurtosis carry out the information
about irregularities characterizing the mass, 
whereas the mean value accounts for
an offset  to be referred to these three parameters.

\subsection{Classification}

The $16$ features extracted from each lesion are classified by a 
 standard three-layer feed-forward neural network with 
$n$ input, $h$ hidden and two output neurons. 
A supervised training based on the back-propagation algorithm with 
sigmoid activation functions both for the hidden
and the output layer has been performed.
The performances of the training algorithm 
were evaluated according to the 5$\times$2 cross validation 
method~\cite{Dietterich}. It is the recommended test to be performed on 
algorithms that can be executed 10 times because it can provide a reliable 
estimate of the variation of the algorithm performances
due to the choice of the training set. 
This method consists in performing 5 replications of the 2-fold cross 
validation method~\cite{Stone}. 
At each replication, the available data are randomly partitioned
into 2 sets ($A_i$ and $B_i$ for $i=1,\dots5$) with an almost equal 
number of entries. 
The learning algorithm is trained on each set and
tested on the other one.
The system performances  are given 
in terms of the sensitivity and specificity values, where
the sensitivity is defined as 
the true positive fraction (fraction of malignant masses correctly classified by the system), whereas the specificity as the true negative 
fraction (fraction of benign masses correctly classified by the system). 
In order to show the trade off
between the sensitivity and the specificity, a 
Receiver Operating Characteristic (ROC) analysis has been 
performed~\cite{Metz,Hanley}. 
The ROC curve is obtained by plotting the  
true positive fraction
versus the false positive fraction 
of the cases (1 - specificity), computed  
while the decision threshold of the classifier is varied.
Each decision threshold results in a corresponding
operating point on the curve.

\section{Image dataset}
\label{sec:dataset}

The image dataset used for this study has been extracted 
from the database of mammograms 
collected in the framework of a collaboration between physicists from 
several Italian Universities and INFN (Istituto Nazionale di Fisica Nucleare) 
Sections, and radiologists from several Italian 
Hospitals~\cite{Bottigli,magic5}. 
The mammograms come both from screening and from the routine work carried out in the participating Hospitals.
The $18 \times 24$ cm$^2$ mammographic films were 
digitized by a CCD linear scanner 
(Linotype Hell, Saphir X-ray), obtaining images characterized by
a  $85 \mu$m pixel pitch and a 12-bit resolution.
The pathological images are fully characterized by a consistent 
description, including the radiological diagnosis, the
histological data and  the coordinates of the center
and the approximate
radius (in pixel units) of a circle drawn by the radiologists around the lesion (truth circle).
Mammograms with no sign of pathology are stored as normal images only  
after a follow up of at least three years.

A set of 226 massive lesions were used in this study: 
109 malignant and 117 benign masses were
extracted from single-view cranio-caudal or lateral  mammograms. 
The dataset we analyzed can be considered as representative of the patient population that is sent for biopsy under the current clinical criteria.

\section{Results}
\label{sec:results}

\begin{figure}[htb]
\vspace{9pt}
\begin{center}
\includegraphics[width=6.5cm]{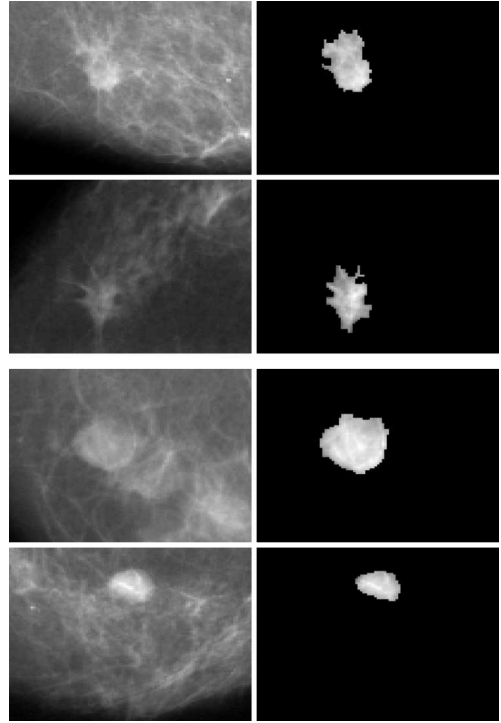}
\end{center}
\caption{\label{fig:segm_examples} 
Examples of segmented masses: two malignant masses (top) and two benign masses (bottom).}
\end{figure}

The 226 massive lesions were segmented by the system and shown to an
 experienced radiologist, whose assistance in accepting or rejecting
 the proposed mass contours was essential for the evaluation of the
 segmentation algorithm efficiency.
Despite the borders of the massive lesions are usually not very sharp in mammographic images, the segmentation procedure we carried out leads to a quite accurate identification of the mass shapes, as can be noticed in fig.~\ref{fig:segm_examples}.
The radiologist 
confirmed only the segmented masses whose contour was sufficiently 
close to that she would have drawn by hand on the image.
The dataset of 226 cases available for our analysis was reduced to 200 
successfully-segmented masses (95 malignant and 105 benign masses),   
thus corresponding to an efficiency $ \epsilon = 88.5\%$ for the 
segmentation algorithm.

Once the 16 features were extracted from each
well-segmented mass, 5 different train and 5 different test sets for the
5$\times$2 cross validation analysis were prepared by randomly
assigning each of the 200 vectors of features to the train or test set
for each of the 5 different trials.
The optimization of the  network performances was obtained  in our case by assigning 3 neurons  to the hidden layer, resulting in a final architecture for the net of  16 input, 3 hidden and 2 output neurons.

\begin{figure}[htb]
\begin{center}
\vspace{9pt}
\includegraphics[width=8.5cm]{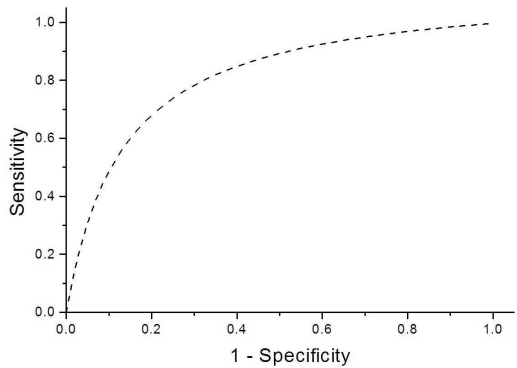}
\caption{ROC curve obtained in the classification of the test set $B_2$ (see tab.~\ref{tab:5X2}).}
\label{fig:ROC}
\end{center}
\end{figure}

The sensitivity and specificity our learning algorithm realized on
each dataset are shown in tab.~\ref{tab:5X2}.
\begin{table}[htb]
\caption{Evaluation of the performances of the standard back-propagation 
learning algorithm for the neural classifier 
according to the 5$\times$2 cross validation method. }
\label{tab:5X2} 
 \begin{tabular}{@{}cccc}  
\hline 
Train Set & Test Set & Sens. (\%) & Spec. (\%)\\
\hline 
$A_1$ & $B_1$ &78.7	&84.9	 \\
$B_1$ & $A_1$ &77.1	&84.6	 \\
$A_2$ & $B_2$ &78.7	&79.3	 \\
$B_2$ & $A_2$ &77.1	&78.9	 \\
$A_3$ & $B_3$ &	80.9	&72.9 \\
$B_3$ & $A_3$ &77.1	&78.7	 \\
$A_4$ & $B_4$ &79.2	&77.4	 \\
$B_4$ & $A_4$ &78.7	&78.9  \\
$A_5$ & $B_5$ &75.6	&81.8	 \\
$B_5$ & $A_5$ &78.0	&74.0	 \\
%
%
\hline 
\end{tabular}
\end{table}
As can be noticed, the performances the neural classifier achieves are
robust, i.e. almost independent of the partitioning of the available
data into the train and test sets.  The average performances achieved
in the testing phase are 78.1\% for the sensitivity and 79.1\% for the
specificity.
The ROC curve obtained in the classification of the test set $B_2$,
containing 100 patterns derived from 47 malignant masses and 53
benign masses, is reported in fig.~\ref{fig:ROC}.  To this curve
belongs the operating point whose values of sensitivity (78.7\%) and
specificity  (79.3\%) are closer to the average values 
achieved on the test sets by the ten different neural networks, 
as shown  in tab.~\ref{tab:5X2}.
As the classifier performances are conveniently evaluated in terms of the area
$A_z$ under the ROC curve, we have estimated this parameter for the
curve plotted in fig.~\ref{fig:ROC}, obtaining $A_z=0.80\pm 0.04$, where
the standard error has been evaluated according to the formula given
by Hanley and McNeil~\cite{Hanley}.

\section{Conclusions and discussion}

The system for the classification of mammographic massive lesions into malignant and benign we realized  aims at improving the radiologist's visual diagnosis of the degree of the lesion malignancy. The system is a semi-automated one, i.e. it segments and analyzes lesions  from physician-located  rectangular ROIs.
As mass segmentation plays a key role in such kind of systems, most efforts have been devoted to the realization  of  a robust segmentation technique. It is  based on edge detection and it works with a comparable efficiency both on malignant and benign masses. The segmentation efficiency  $ \epsilon = 88.5\%$ has been evaluated with the assistance of an experienced radiologist who accepted or rejected  each segmented mass. 
With respect to a number of automated or semi-automated systems with a similar purpose and using a similar approach  already discussed in the literature~\cite{Wirth,Amini,Sahiner,Matthew}, 
the system we present is characterized by 
a robust segmentation technique: 
it is based on an edge-detection  algorithm completely free  
from any application-dependent parameter.

Sixteen  features based on shape, size and intensity  have been extracted out of each  segmented area and merged by a neural decision-making system. 
The neural network performances have been  evaluated in terms of the ROC analysis, obtaining an estimated area under the ROC curve   $A_z=0.80\pm 0.04$.
It gives the indication that the segmentation procedure we developed provides a quite accurate approximation of the mass shapes and that the features we took into account for the classification have a good discriminating power.

\section*{Acknowledgments}

We are grateful to the professors and the radiologists of the collaborating Radiological Departments for  their medical support during 
the database acquisition.
We acknowledge Dr S.~Franz from ICTP (Trieste, Italy) for useful discussions.  Special thanks to Dr M.~Tonutti from Cattinara Hospital (Trieste, Italy) for her essential contribution 
to the present analysis.

\end{document}